\documentclass[aps,pra,nofootinbib,twocolumn]{revtex4}

\catcode`@=11     
\def\eqalign#1{\null\,\vcenter{\openup\jot\m@th
  \ialign{\strut\hfil$\displaystyle{##}$&$\displaystyle{{}##}$\hfil
      \crcr#1\crcr}}\,}
\newcommand{\beq}{\begin{equation}}
\newcommand{\eeq}{\end{equation}}
\newcommand{\four} {  {}^{(4)}\kern-1pt  }
\newcommand{\ben}{\begin{eqnarray}}
\newcommand{\een}{\end{eqnarray}}
\newcommand{\f}{\frac}
\newcommand{\nn}{\nonumber}
\newcommand{\ds}{\displaystyle}
\catcode`@=12   
\begin{document}

\title{\begin{flushright}
gr-qc/0207076 \\
$~$ \\
\end{flushright}
Group of boost and rotation transformations\\
with two observer-independent scales}

\author{{\bf Nicola Rossano~BRUNO}}
\affiliation{Dipartimento di Fisica, Universit\`a di Roma ROMA TRE,\\Via Vasca Navale 84, 00146 Roma, Italy\\I.N.F.N. - Sezione di ROMA TRE, Roma, Italy}

\begin{abstract}
I examine the structure of the deformed Lorentz transformations
in one of the recently-proposed schemes
with two observer-independent scales.
I develop a technique for the analysis of general combinations
of rotations and deformed boosts.
In particular, I verify explicitly that the transformations form group.
\end{abstract}
\maketitle


\section{Introduction}
Recently many modern approaches to the problem of unification of
Quantum Mechanics and General Relativity have led to arguments
that suggests a modification of Lorentz symmetry at the Planck Scale.
As a possible tool for this ongoing quantum-gravity debate
Amelino-Camelia proposed~\cite{Amelino-Camelia:2000mn,Amelino-Camelia:2000ge}
the possibility that the rotation/boost transformations between
inertial observers might be characterized
by two rather than one observer-independent
scale: in addition to the velocity scale, $c$, one introduces
a length (momentum) scale $\lambda$ ($1/\lambda$).
In Refs.~\cite{Amelino-Camelia:2000mn,Amelino-Camelia:2000ge}
an illustrative example of the new type of transformation laws
was also analyzed, obtaining results, including the emergence
of a maximum momentum $1/\lambda$, in leading order in
the second observer-independent scale $\lambda$.
In Ref.~\cite{Bruno:primo} the analysis of that model was
generalized to all orders in $\lambda$, and in particular it was verified
that indeed the new transformation rules saturate
at maximum momentum $1/\lambda$.

In these past few months interest in this idea has increased as various
authors~\cite{MagueijoSmolin,dafr,KowalskiNowak,LukierskiNowickiDSR,JudesVisser}
considered different ways to introduce the two observer-independent
scales, still following the general scheme proposed
in Refs.~\cite{Amelino-Camelia:2000mn,Amelino-Camelia:2000ge}.
These relativistic theories are being
called ``Doubly Special Relativity" (DSR),
and their possible relevance for the study of noncommutative
spacetimes and loop quantum gravity is under
investigation~\cite{Amelino-Camelia:2000mn,MagueijoSmolin,KowalskiNowak}.
In the physics of quantum spacetime and quantum gravity DSR may
provide a tool for the description of the Planck length as a kinematical
scale of the structure of spacetime or energy-momentum space,
upon identification of $\lambda$ with the Planck length: $\lambda \sim L_p$.
The observable implications of DSR theories are being studied
mostly in relation with forthcoming powerful Lorentz-symmetry
tests~\cite{grbgac,glast} and in searches of a
kinematical solution for the puzzling observations
of ultra-high-energy cosmic rays~\cite{cosmicRAYdata,kifune,AmelinoPiranPRD}.

Also under
investigation~\cite{Amelino-Camelia:2000mn,KowalskiNowak,LukierskiNowickiDSR}
is the role that $\kappa$-Poincar\'{e} Hopf
algebras~\cite{LukierskiRuegg1992,Majid:1994cy,lukbasis}
can play in these DSR theories.
In the one-particle sector of the
DSR theories so far considered $\kappa$-Poincar\'{e} mathematics
has a role which is analogous to the role of Lorentz/Poincar\'{e} mathematics
in Einstein's Special Relativity.
For the two-particle sector it is still unclear if and in which
way $\kappa$-Poincar\'{e} could have a role.

The connection with $\kappa$-Poincar\'{e} may be cause of concern
with respect to the group properties of the DSR transformations;
in fact, for the Lorentz sector of $\kappa$-Poincar\'{e} Hopf
algebras it is generally expected~\cite{lukQUASIgroup}
that the transformations obtained by exponentiation of the
boost/rotation generators would not form group: they would
only form quasigroup in the sense of Batalin~\cite{batalin},
As experimentalists are preparing~\cite{glast} for investigations that have the sensitivity to test Planck-scale modifications of Lorentz symmetry, it is important to settle these possible concerns about the group structure, since they may affect the willingness of experimentalists to include DSR predictions among the tests they actually perform. If DSR symmetries only lead to quasi-group structure, as generally expected for theories based on $\kappa$-Poincar\'e mathematics, one might conclude that these symmetries are not suitable for the description of Planck-scale physics. In fact  the structure coefficients of a quasi-group depend on the initial variables on which the transformations act. To characterize the puzzling implications of this dependence one can consider three inertial observers that are performing measures on the four-momentum of a particle. Quasi-group structure implies \cite{batalin} that there must exist a unique transformation that relates the measures performed by the first and the third inertial observer, but this transformation depends on the momenta observed by the second inertial observer.

With respect to this issue I here reconsider the theory ``DSR1",
the one used as illustrative example in
Refs.~\cite{Amelino-Camelia:2000mn,Amelino-Camelia:2000ge}.
In Ref.~\cite{Amelino-Camelia:2000mn} it was argued that
the specific $\kappa$-Poincar\'{e} Hopf
algebra that is relevant for DSR1 is a special case in which
the exponentiation of the generators in the Lorentz sector
does lead to an ordinary symmetry group.
The observation in Ref.~\cite{Amelino-Camelia:2000mn}
was based on the fact that the relevant $\kappa$-Poincar\'{e} Hopf
algebra has a classical (undeformed) Lorentz sector:
the commutation relations for rotation/boost generators are
just the usual Lorentz ones, although the differential representation
of the boost generators on energy-momentum is deformed (nonlinear).
This in particular implies that the Batalin conditions~\cite{batalin}
for group structure\footnote{Batalin basically observed~\cite{batalin}
that an ordinary group will be obtained by exponentiation of
the generators if the algebra truly closes on its generators.
In generic $\kappa$-Poincar\'{e} Hopf
algebras some commutators of the generators of the Lorentz sector
are expressed in terms of translation generators, and this causes the
transformations obtained by exponentiation of the Lorentz-sector
generators to form only quasigroup rather than
group~\cite{lukQUASIgroup,batalin}.}
are satisfied.
Since the algebra is the same, using the Baker-Campbell-Hausdorff formula
one deduces~\cite{Amelino-Camelia:2000mn} that by exponentiation of the
DSR1 rotation/boost generators one obtains genuine group elements,
which actually combine just as usual: one obtains again the Lorentz
group, only realized nonlinearly. 
 
Following the same line of analysis one can show that the other versions of $\kappa$-Poincar\'{e} Hopf
algebras considered in the past form only quasi-group with the mentioned problem of physical interpretation.

However this argument based on BCH is not enough to establish that the path from the algebra to the group properties can be legitimately followed. There are some mathematical issues to be investigated, including the assumption that the BCH series converges (see {\it e. g.} Ref.~\cite{bourbaki} and the references therein). Rather than give these proofs I here settle the issue through an explicit verification of the group properties of the DSR1 transformations. In addressing this issue I also develop a technique for the analysis of DSR1 deformed boost/rotation transformations which should
prove useful in applications of the formalism; in fact, the nonlinear
structure of the DSR1 transformation rules of course prevents us
from describing boosts through matrices and the calculations
can in some cases become rather involved. The technique I introduce
is useful in handling these difficulties.

In section \textsc{ii} I revise some issues relevant for the
Baker-Campbell-Hausdorff (BCH) formula that emerge naturally
in the composition of two exponential of operators that do not
commute. Notice that in ordinary Special Relativity
we do not need to resort to BCH because the Lorentz generators
have a simple matrix representation.
The group laws in that case simply are the
matrix products while in the DSR1-deformed case we do not have
a matrix representation and we have to directly deal with BCH.

In section \textsc{iii} I introduce a system of  differential equations
whose solutions are the deformed Lorentz transformations.
To derive them I exploit a formalism~\cite{feynman}
introduced by Feynman in 1951
as a tool for the analysis of QED perturbation theory.

In section \textsc{iv} I consider some explicit applications
of the formulas obtained in Section \textsc{iii}, with emphasis on the
emergence of the expected group properties.
Section \textsc{v} presents some closing remarks.

\section{Group properties and Baker-Campbell-Hausdorff formula}
The boost and rotation generators adopted in the DSR1 scheme
are~\cite{Amelino-Camelia:2000mn,Bruno:primo}:
\ben
N_i\!\! &=& \!\! ik_i\f{\partial}{\partial \omega}
\!\!+\!i\left( \f{\lambda}{2} \vec{k}^2
\!\!+\! \f{1-e^{-2\lambda\omega}}{2\lambda}\right )
\f{\partial}{\partial k_i}\!\!-i\lambda k_i \left(k_j\f{\partial}{\partial k_j}
\right) ;\nn\\
M_i\!\! &=&\!\! -i\epsilon_{ijk} k_j \f{\partial}{\partial k_k}
\label{action}
\een
These generators can be naturally described as part of the so-called
bicross-product-basis $\kappa$-Poincar\'{e} Hopf
algebra~\cite{Majid:1994cy,lukbasis}, which in the algebra
sector\footnote{The coalgebra sector,
which plays no explicit role in my analysis, is described in
Refs.~\cite{Majid:1994cy,lukbasis}.} prescribes the commutation rules
\ben
[M_{\mu\nu},M_{\rho\tau}]\!\!&=&\!\! i\left (
\eta_{\mu\tau}M_{\nu\rho}\!-\!\eta_{\nu\rho}M_{\nu\tau}
\!+\!\eta_{\nu\rho}M_{\mu\tau}\!-\!\eta_{\nu\tau}M_{\mu\rho} \right )\,;\nn\\
 {[} M_{i}, k_{j} ] \!\!&=&\!\!
i {\epsilon}_{ijk} k_k \,; \,\,\,\,\,\,\,[M_i,\omega ]=0\,; \nn\\
 {[}N_i,k_j] \!\!&=&\!\! i\delta_{ij}\left (
\f{1}{2\lambda}\left(1-e^{-2\lambda\omega}\right )
+\f{\lambda}{2}\vec{k}^2\right )-i\lambda k_i k_j\, ;\nn\\
 {[}N_i,\omega ]\!\!&=&\!\!ik_i\, ;\nn\\
 {[}P_\mu,P_\nu]\!\!&=&\!\!0\label{algebrasector}
\een
where $P_\mu =(\omega,\vec{k})$ are the translation generators
and $M_{\mu\nu}$ are the Lorentz-sector generators with rotations
given by $M_{k}=\f{1}{2} \epsilon_{ijk} M_{ij}$ and
boosts $N_i=M_{0i}$.


The special feature of the
bicross-product-basis is the fact that
the Lorentz-sector commutators are undeformed,
only the commutators between boosts and
translation generators are deformed.
This fact has important implications
for the group properties of finite transformations.

The DSR1 finite rotation/boost transformations on
energy-momentum space are
constructed~\cite{Amelino-Camelia:2000mn} by the action
of the exponentiation of the generators.
This action was first
studied~\cite{Amelino-Camelia:2000mn,Amelino-Camelia:2000ge}
only in leading order in $\lambda$, and then,
in Ref.~\cite{Bruno:primo}, the exact (all orders in $\lambda$)
form of the transformations was analyzed for the special case of
the action on energy-momentum space
of the exponentiation of one boost generator.
Here I consider general rotation/boost transformations
of the form
\beq
P_\mu =  e^O P_\mu^0  e^{-O} \label{LTs}
\eeq
where $O\!\!=\!-i\sum_i \xi_i O_i$ and $O_i\!\!=\!\!\{ N_1, N_2, N_3, M_1,M_2,M_3 \}$.

For the composition of two such transformations
one can naturally~\cite{Amelino-Camelia:2000mn,Amelino-Camelia:2000ge}
resort to the Baker-Campbell-Hausdorff's formula (BCH)
\ben
H(A,B)&=& A+B+\f{1}{2} [A,B]+\f{1}{12}[A,[A,B]]+\nn\\
&+&\f{1}{12}[B,[B,A]]-\f{1}{24}[A,[B,[A,B]]]+....\label{Haus}
\een
where $A$ and $B$ are two operators with known
commutation relations.

It is useful to consider a generic element of a
Lie algebra $g\,\in $ {\large G} such
that  $g=\sum_i \alpha_i g_i$ where $\ds \{ g_i \}$ are
the generators of {\large G} and  $\alpha_i$ complex numbers.
One finds \cite{bourbaki} that {\large G} is a group with product  $H({\cdot},{\cdot})$.
In fact BCH introduces only commutators between algebra generators
that are closed in the algebra by definition (\ref{algebrasector}),
and then $H({\cdot},{\cdot})$ is a linear combination of generators with
coefficients that only depend on the $\alpha_i$ and on the structure
coefficients.
Exploiting this fact one can easily
prove that the transformations done by the  exponentiation of a
Lie algebra do form group with product done by multiplication of
exponentials. The identity obviously is $\ds e^0=1$, and for each
element of the group  $\ds e^g$ there is an inverse $\ds e^{-g}$
such that $\ds e^g\, e^{-g}=1$.

For the DSR1 theory, since the action on momentum
space is nonlinear,
we do not have a matrix representation of boosts,
so this BCH argument is important in arguing
that arbitrary rotation and boost transformations
form group.
However there are some mathematical
issues~\cite{bourbaki}
which may require consideration.
I shall verify here explicitly the group properties.

\section{General DSR1 transformations}
Equation (\ref{LTs}) governs the relation between the
energy-momentum $(\omega^0 , \vec{k}^0)$ attributed to a particle
by a given inertial observer
and the energy-momentum $(\omega , \vec{k})$
attributed to that same particle
by another inertial observer.
In order to analyze the dependence of $(\omega , \vec{k})$
on the six parameters $\xi_i$ that appear in the description
of the general rotation/boost transformation with generator $O$
we need to consider the derivatives $\ds \f{dO}{d\xi_i}$ that are done by
Feynman formalism
\beq
\f{d}{d\xi_i}e^{O}=\int_0^1 e^{(1-s)O}\, \f{dO}{d\xi_i} e^{sO} ds
\eeq
Taking into account this rule of derivation\footnote{Shortly speaking
if we append an index $s$ to two noncommuting operators such
that $A_sB_{s'}$ is equal to $AB$ if $s>s'$ and $BA$
if $s<s'$ we can equally write $A_sB_{s'}$ or $B_sA_{s'}$.
The index $s$ should not to be necessarily a discrete one.
In particular we can substitute 
$A$ with $\int_0^1 A_s ds$ because for
only one operator $s$ is unnecessary and $A\int_0^1 ds=A$,
but with the advantage that
can be considered like an ordinary function. In our case
we can replace $O=\int_0^1 O_s ds$. The derivative now is an
ordinary derivative of a
function, {\it i.e.} $\ds \f{d}{d\xi}e^{\int_0^1 O_s ds}
=\int_0^1 e^{\int_0^1 O_{s'} ds'} O_s ds$.
Sharing out the integration in $ds'$ such 
that $\int_0^1 O_{s'}ds'=\int_s^1O_{s'}ds'+
\int_0^s O_{s'}ds'$ one gets $\ds \f{d}{d\xi}e^{O}
=\int_0^1 e^{\int_s^1O_{s'}ds'}\left (\f{dO}{d\xi}
\right )_{\!\!\! s} e^{\int_0^s O_{s'}ds'} ds
=\int_0^1 e^{(1-s)O}\f{dO}{d\xi}e^{sO} ds$
where I eliminated the indexes because the operators were
all in the right order. For a more exhaustive tractation
see the original paper~\cite{feynman}.}
 one obtains
\ben
\f{dP_\mu}{d\xi_i}&=&\left (\int_0^1 e^{(1-s)O}\, \f{dO}{d\xi_i} e^{-(1-s)O}
ds\right ) P_\mu + \nn\\
&-&P_\mu \left (\int_0^1 e^{sO}\, \f{dO}{d\xi_i} e^{-sO} ds\right )
\een
The integrals in $ds$ can be performed by means of Sophus Lie's expansion
\ben
e^{tA}Be^{-tA}&=&\sum_0^\infty \f{t^n}{n!}C_n(A,B)\nn\\
C_{n+1}(A,B)&=&\left [A,C_n(A,B)\right ]\,\,\,\,\,\,\,\,\,C_0(A,B)=B
\een
obtaining
\ben
\f{dP_\mu}{d\xi_i}&=& \sum_{n=0}^\infty \f{1}{(n+1)!} \left [
C_n \left (O,\f{d O}{d\xi_i} \right ),P_\mu \right ]=\nn\\
&=& \left [ \f{d O}{d\xi_i}, P_\mu \right ] + \f{1}{2}\left [
\left [O, \f{d O}{d\xi_i}\right ], P_\mu \right  ]+.... \label{variation}
\een
Making use of the algebra commutators, one can rewrite the
right-hand side in the form
\beq
\f{dP_\mu}{d\xi_i}=-i\sum_{j=1}^6 \left [
\beta(\{ \xi \})_i^j O_j,P_\mu \right ]\label{recursion}
\eeq
These formulas are the key to the full description of
general rotation and boost transformations in DSR1.
As shown in the following in some explicit examples,
the coefficients $\ds \beta(\{ \xi \})_i^j$ can be found
by recursion formulas.

\section{Detailed analysis of some examples of DSR1 transformations}
In this section I illustrate the application of
the formalism introduced in the preceding section,
and, in particular, I verify
explicitly the key requirements for group structure of
DSR1 transformations.

\subsection{Composition of two Lorentz transformations along the
same direction}
Rotations are undeformed in DSR1, so they pose no difficulty.
The simplest nontrivial case is the composition of two boosts
along the same direction.
Let me denote that direction with ``1'', so that the relevant boost
generator is $N_1$,
and consider a boost
with rapidity $\xi_1$ and another boost with
rapidity $\xi_2$.
The two transformations in succession
yield $\ds P_\mu = e^{-i\xi_1 N_1}
e^{-i\xi_2 N_1}P^0_\mu e^{i\xi_1 N_1}e^{i\xi_2 N_1}
= e^{-i(\xi_1 +\xi_2) N_1}P^0_\mu e^{i(\xi_1 +\xi_2)N_1} $
that is again a boost along direction ``1'' with
rapidity $\xi_1+\xi_2$. In this case the application
of (\ref{variation}) is very easy
because 
only the first commutator does not vanish.
In this way we get the eight differential equations
\beq
\f{dP_\mu}{d\xi_1}=-i(N_1P_\mu);\,\,\,\,\,\,\,\, \f{dP_\mu}{d\xi_2}
=-i(N_2P_\mu)
\label{onedirection}
\eeq
which can be analyzed with the same techniques introduced
in Ref.~\cite{Bruno:primo}. In this paper I use a technique which can be used in alternative to the one of \cite{Bruno:primo} and in most cases give a better chance of finding explicit solutions. The natural way to illustrate this technique is in the simple context considered in this subsection\footnote{Of course, in the simple case of the composition of two boosts along the same direction, here considered only for illustrative purposes, this technique is not necessary (one could use the technique introduced in \cite{Bruno:primo}).}.

Suppose that we know how to construct the transformations done
by two boosts in succession. If they are equal to the unique
transformation whose differential equations are done
by (\ref{onedirection}) then they must satisfy (\ref{onedirection}).
In this case the two boosts in succession must be equivalent to the
single boost transformation with rapidity $\xi_1+\xi_2$.
The end result of
two transformations in succession is easily analyzed knowing
the action of a single boost \cite{Bruno:primo}. Consider a boost transformation in
direction ``1'' $\ds P^{'}_\mu =P^{'}_\mu (\xi_2;P_\mu^0)$ followed by
a second boost transformation in the same direction  $\ds P_\mu
=P_\mu (\xi_1;P_\mu^{'})$
we then clearly  have
\beq
P_\mu \!\!=\!\! P_\mu\! (\xi_1,\xi_2;P_\mu^0)
\!\!=\!\!\! P_\mu\! (P_\mu^{'}\! (\xi_2;P_\mu^0),\xi_1)
\!\!=\!\!\!P_\mu\! (\xi_1\!\! +\!\!\xi_2;P_\mu^0)
\label{composition}
\eeq
Substituting them
with the (\ref{onedirection}) (with the action of generators
specified by (\ref{action})) one gets an identity. 
This illustrates the strategy of application of formula (\ref{variation}).

\subsection{Composition of two boost transformations along
different directions}
Now I consider a somewhat more complicated case:
composition of two boosts along
different direction. Let me denote with ``2'' the direction of one of
the boosts, with
rapidity $\xi_2$, and with ``1'' the direction of the other
boost, with
rapidity $\xi_1$. This case involves the generators $N_1$, $N_2$
that do not commute and all terms in (\ref{variation}) do not vanish.
In particular to express the composition of these two boosts in terms
of a single
Lorentz transformation it is necessary to calculate
the coefficients $a,\, b,\, c$ of the BCH $\ds P_\mu
=e^{-i\xi_1 N_1}e^{-i\xi_2 N_2}P_\mu^0 e^{i\xi_2 N_2}e^{i\xi_1 N_1}
=e^{-i(a N_1+b N_2+c M_3)}P^0_\mu e^{i(a N_1+b N_2+c M_3)} $
where $M_3$ is proportional to $[N_1,N_2]$, $i.e.$ $M_3$ is the generator
of rotations around direction ``3''. From the knowledge
of  $a,\, b,\, c$ the $\beta_i^j$ coefficients in (\ref{recursion})
can be determined by recursion. In fact, calculating the commutators
in (\ref{variation}) one only encounters the
operators $O_1=N_1,\, O_2=N_2,\, O_3=M_3$
and (\ref{variation}) can be written as
$\ds  \f{dP_\mu}{d\xi_i}
=-i\sum_{n=0}^{\infty}\left ( (\beta^1_i)^n N_1
+ (\beta^2_i)^n N_2 + (\beta^3_i)^n M_3\right )$.
The coefficients at each order $n$ are given by
\ben
(\beta^1_i)^n=\f{1}{(n+1)!}[c(\beta^2_i)^{n-1}-b(\beta^3_i)^{n-1} ]\nn\\
(\beta^2_i)^n=\f{1}{(n+1)!}[a(\beta^3_i)^{n-1}-c(\beta^1_i)^{n-1} ]\nn\\
(\beta^3_i)^n=\f{1}{(n+1)!}[a(\beta^2_i)^{n-1}-b(\beta^1_i)^{n-1} ]
\label{explrecursion}
\een
where $\ds (\beta^1_i)^0=\f{da}{d\xi_i}$, $\ds (\beta^2_i)^0
=\f{db}{d\xi_i}$, $\ds (\beta^3_i)^0=\f{dc}{d\xi_i}$.
Exploiting the fact that the Lorentz algebra is unmodified in DSR
the coefficients $a,\, b,\, c$ can be determined
using matrix representation of the algebra generators
(the DSR deformation will then be manifest in the fact that those same
coefficients enter different formulas, which take into account
the nonlinear realization of the Lorentz algebra\footnote{
This observation is in agreement with similar results obtained in \cite{LukierskiNowickiDSR} and will be further confirmed by equs. (\ref{variationort}).
}
).
Let me start with the matrix representation of the Lorentz generators.
They can be constructed expanding the corresponding exponentials.
For example consider a boost in direction ``1''
\ben
&P_\mu & =\!\![\exp (-i\xi_1 N_1)]_{\mu\nu}(P^0)^\nu
=\!\!\left[ \sum_{n=0}^\infty \f{(-i\xi_1 N_1)^n}{n!}\right]_{\mu\nu}P_0^\nu=
\nn\\
&=&\!\!\!\!\!\!\left(\!\!
\begin{array}{cccc}
\cosh\xi_1\!\! & \sinh\xi_1\!\! & 0\!\! & 0 \\
\sinh\xi_1\!\! & \cosh\xi_1\!\! & 0\!\! & 0 \\
0\!\! & 0\!\! & 1\!\! & 0 \\
0\!\! & 0\!\! & 0\!\! & 1
\end{array}
\!\!\right )
\!\!\!\!\left (\!\!
\begin{array}{c}
\omega^0 \\ k_1^0 \\ k_2^0 \\ k_3^0
\end{array}
\!\!\right)\!\!=\!\!
\exp\!\!\!{\left [ \!-i\xi_1\!\!\left(\!\!
\begin{array}{cccc}
0\!\! & i\!\! & 0\!\! & 0 \\
i\!\! & 0\!\! & 0\!\! & 0 \\
0\!\! & 0\!\! & 0\!\! & 0 \\
0\!\! & 0\!\! & 0\!\! & 0
\end{array}
\!\!\right )\!\!\right ] }
\!\left (\!\!
\begin{array}{c}
\omega^0 \\ k_1^0 \\ k_2^0 \\ k_3^0
\end{array}
\!\!\right)\nn\\
\een
where 
the right side has been obtained solving
the differential equations that are associated to
this boost transformation. 
In the same manner one gets the matrix form of the other generators
here of interest $N_2$, $M_3$.

The next step is to get the unique
transformation $\ds e^{-i(aN_1+bN_2+cM_3)}$
yielded by $\ds e^{-i\xi_1N_1}e^{-i\xi_2N_2}$.
I have to determine the three
coefficients $a,\,b,\, c$ as
functions of $\xi_1,\,\xi_2$.
To perform this task I can first calculate
these transformation in matrix form,
the first in terms of $a,\,b,\, c$,
the second in terms of $\xi_1,\,\xi_2$,
and then comparing three matrix elements.
Following this line of analysis I get a non-linear
system of three coupled equations
\ben
\f{(a^2+b^2)\cosh d - c^2}{d^2}&=&\cosh\xi_1 \cosh\xi_2\nn\\
\f{(a^2-c^2)\cosh d + b^2}{d^2}&=&\cosh\xi_1 \nn\\
\f{(b^2-c^2)\cosh d + a^2}{d^2}&=&\cosh\xi_2 \label{compare}
\een
where $d^2=a^2+b^2-c^2$. A solution of (\ref{compare})
is given by
\ben
a&=&\sqrt{\cosh\xi_1-1}\sqrt{\cosh\xi_2+1}\,\,\alpha (\xi_1,\xi_2)\nn\\
b&=&\sqrt{\cosh\xi_2-1}\sqrt{\cosh\xi_1+1}\,\,\alpha (\xi_1,\xi_2)\nn\\
c&=&\sqrt{\cosh\xi_1-1}\sqrt{\cosh\xi_2+1}\,\,\alpha (\xi_1,\xi_2)\nn\\\nn\\
\alpha (\xi_1,\xi_2)\!\!\!&=&\!\!\!\f{\mbox{arccosh}\left [\f{1}{2}(\cosh\xi_1
\!\!\cosh\xi_2\!\!+\!\!\cosh\xi_2\!\!-\!\!1)
\right ]}{\sqrt{\cosh\xi_1 \!\!\cosh\xi_2\!\!+\!\!\cosh\xi_2\!\!+\!\!\cosh\xi_1\!\!-\!\!3}}\label{BCHcoeff}
\een
One can see that $a(\xi_1=0,\xi_2)=0$, $a(\xi_1,\xi_2=0)=\xi_1$
and analogous relations hold for $b,\, c$ as expected.
Exploiting equation (\ref{variation}) and recursion
formulas (\ref{explrecursion}) one finds in this case
\ben
\f{dP_\mu}{d\xi_1}\!\!=\!\!-i(N_1 \! P_\mu);\,\,\f{dP_\mu}{d\xi_2}
\!\!=\!\!-i(\cosh\!\xi_1 \! N_2\!\!+\!\!\sinh\!\xi_1 \! M_3 )P_\mu
\label{variationort}
\een
Note that these equations are the same that one could find
in a purely classical framework. To show that deformed
Lorentz transformations satisfy (\ref{variationort}) 
one must construct (following the same technique that
we used for the previous example) the complete transformations
yielded by two boosts in succession
\beq
P_\mu = P_\mu (\xi_1,\xi_2;P_\mu^0)
=P_\mu(P_\mu^{'} (\xi_2;P_\mu^0),\xi_1)
\label{compositionort}
\eeq
from Ref. \cite{Bruno:primo}.
It is easy to show that putting (\ref{compositionort})
in (\ref{variationort}) we get again an identity. As in the previous example this shows explicitly one of the group properties of deformed Lorentz transformations.

Exploiting the same technique for the composition of two arbitrary deformed Lorentz transformations one can verify all group properties.

\subsection{On arbitrary boost transformations}
As a third application of the formula (\ref{variation}) I now consider
an arbitrary Lorentz transformation given by exponentiation of
the operator $O=-i(\xi_1 N_1 + \xi_2 N_2)$. 
In this case equation (\ref{variation}) reads
\beq
\f{dP_\mu}{d\xi_1}=-i(A(\xi_1,\xi_2)N_1-B(\xi_1,\xi_2)N_2
-C(\xi_1,\xi_2)M_3) P_\mu
\eeq
where
\ben
A&=&1+\f{\xi_2^2}{3!}+\f{\xi_1^2\xi_2^2}{5!}
+\f{\xi_2^4}{5!}+\f{\xi_1^4\xi_2^2}{7!}+2\f{\xi_2^4\xi_1^2}{7!}
+\f{\xi_2^6}{7!}+....\nn\\
B&=&\f{\xi_1\xi_2}{3!}
+\f{\xi_1^3\xi_2}{5!}+\f{\xi_1\xi_2^3}{5!}+\f{\xi_1^5\xi_2}{7!}
+2\f{\xi_2^3\xi_1^3}{7!}+\f{\xi_1\xi_2^5}{7!}+....\nn\\
C&=&\f{\xi_2}{2!}+\f{\xi_1^2\xi_2}{4!}+\f{\xi_2^3}{4!}
+\f{\xi_1^4\xi_2}{6!}+2\f{\xi_1^2\xi_2^3}{6!}+\f{\xi_2^5}{6!}+....
\een
In this case it is more difficult to derive the exact form of $A$, $B$, $C$
from the terms known  order by order.
However, it is clear that the series
converges because the following relations hold
\ben
A \leq  \cosh\xi_1 \cosh\xi_2 &;& \,\,\,\,|B|\leq |\sinh\xi_1\sinh\xi_2|;\nn\\ |
C| &\leq & |\sinh\xi_2\cosh\xi_1|
\een
for each value of $\xi_1,~\xi_2$.
In the same manner we can analyze $\ds\f{dP_\mu}{d\xi_2}$
with analogous results.

In this subsection,
my analysis gave
another illustration of the usefulness
of (\ref{variation}) as a useful tool
for calculations. In more complicated cases, such as the one
considered in this subsection, the formula (\ref{variation})
is still useful, although close-form all-order results are
troublesome. 
I want to point out that an all-order result on arbitrary boost transformations has been obtained, following an alternative approach (not differential but algebraic), in Ref. \cite{LukierskiNowickiDSR}.

\section{Closing remarks}
The analysis reported here completes a natural set of
consistency checks for the structure of DSR1 in the energy-momentum
one-particle sector. DSR1 transformations clearly form a genuine group
of symmetries of a plausible energy-momentum space, which
is in full agreement with our presently-available low-energy
data but predicts observably large new effects for forthcoming
experiments~\cite{Amelino-Camelia:2000mn,glast}. 

This robust starting point should provide additional motivation
for the study of the two-particle sector, where
some issues still require further
study~\cite{Amelino-Camelia:2000mn,dafr,JudesVisser},
and for a corresponding spacetime
realization~\cite{Amelino-Camelia:2000mn,KowalskiNowak}
of DSR symmetries.

Notice that the same BCH argument developed in section \textsc{ii} can be applied in general to all DSR theories presented, and in particular to Magueijo-Smolin DSR2~\cite{MagueijoSmolin}. This suggests that the techniques here proposed for DSR1 might be easily adapted to DSR2 and possible other DSR proposals.

\section*{Acknowledgments}
I am grateful to Giovanni Amelino-Camelia for suggesting
this research project and for numerous discussions on
the development of my studies as part of his
supervision of my Laurea thesis.


\end{document}